\documentclass[USenglish,twocolumn]{article}

\usepackage[utf8]{inputenc}
\usepackage[numbers,sort&compress]{natbib}
\usepackage{booktabs}
\usepackage{tabularx}
\usepackage{array}
\usepackage{tikz}
\usetikzlibrary{arrows.meta,positioning}
\usepackage[big]{dgruyter}

\newcolumntype{Y}{>{\RaggedRight\arraybackslash}X}
\newcommand{\code}[1]{\texttt{#1}}
\newcommand{\tool}[1]{\emph{#1}}

\newcommand{\Rag}{Retrieval-augmented generation}

\begin{document}

\articletype{Research Article{\hfill}Open Access}

\author*[1]{Marek Horv\'ath}
\author[2]{Em\'ilia Pietrikov\'a}

\affil[1]{Technical University of Ko\v{s}ice, Faculty of Electrical Engineering and Informatics, Department of Computers and Informatics, Ko\v{s}ice, Slovakia, E-mail: marek.horvath@tuke.sk}
\affil[2]{Technical University of Ko\v{s}ice, Faculty of Electrical Engineering and Informatics, Department of Computers and Informatics, Ko\v{s}ice, Slovakia, E-mail: emilia.pietrikova@tuke.sk}

\title{\huge Evaluating Semantic and Quality-Aware Retrieval for Source Code Repositories}
\runningtitle{Semantic and quality-aware retrieval for source code}

\begin{abstract}
{Keyword-based retrieval is limited for source-code repositories when queries are expressed in natural language or concern implementation intent and code quality rather than exact tokens. This study evaluates a prototype retrieval system that combines function-level fragmentation, text-and-code embeddings, \tool{ChromaDB} vector storage, LLM-derived quality metadata, and four retrieval modes: semantic, quality-filtered, hybrid, and automatic routing. The concrete evaluation uses an educational C-code corpus. The full corpus contains 563 anonymized programmer identifiers and 8,951 C files; a reproducible 10\% indexed sample contains 56 programmer identifiers, 847 files, and 3,839 fragments. Across 15 manually judged queries, semantic retrieval achieved nDCG@5 of 0.820, Success@5 of 0.800, and MRR of 0.644. The automatic router selected the expected mode for all 15 queries. In a small manual audit, LLM-derived quality scores were within one point of the manual assessment for 9 of 12 fragments. Within the reported query set, semantic retrieval was the strongest overall mode, while explicit quality metadata was most useful for explicitly quality-oriented queries.}
\end{abstract}


\journalname{Open Computer Science}
\journalyear{2026}
\startpage{1}

\maketitle

\section{Introduction}

Source-code repositories are difficult to inspect when the search intent concerns behavior, design choices, or code quality rather than exact tokens. Exact-token search remains useful when the user knows a function name, library call, or identifier, but it is less suitable for intent-level questions such as which fragments read from files, use dynamic memory allocation, or handle input errors. Such questions require retrieval based on behavior and context rather than literal token overlap. Related source-code exploration work has also used static analysis and programmer profiles to summarize code artifacts beyond direct keyword matching \cite{pietrikova2015profile}.

Natural-language code search addresses part of this problem by mapping queries and code fragments into a shared vector space \cite{Neelakantan2022}. In source-code repositories, however, topical similarity is not the only useful signal. Users may also ask quality-oriented questions about readability, error handling, input validation, or memory management. Adding explicit quality metadata can support such questions, but it can also distort topical relevance by promoting fragments that score well on quality dimensions while only weakly matching the requested functionality.

Educational C-code repositories provide a useful concrete evaluation setting because many implementations solve related tasks under comparable constraints. Such repositories occur in programming courses that combine repeated assignments, automated assessment, and game-creative or problem-based learning activities \cite{pietrikova2015towards,pietrikova2025game}. At the same time, findings from such a corpus should not be generalized to other repository types without further evaluation. This study therefore compares retrieval modes rather than assuming that quality scoring necessarily improves search. The evaluated prototype indexes source-code fragments, attaches LLM-derived quality metadata, stores embeddings and metadata in \tool{ChromaDB}, and supports semantic, quality-filtered, hybrid, and automatically routed retrieval. Retrieved fragments may also be passed to a citation-grounded answer-generation step, but the evaluation reported here focuses on retrieval and supporting components.

The study reports an empirical prototype evaluation for semantic and quality-aware retrieval over source-code fragments. It uses an educational C-code corpus as the concrete evaluation setting, together with a fixed indexed sample, query set, relevance judgments, and recorded evaluation outputs. The contributions are:
\begin{itemize}
  \item a reproducible pipeline for function-level semantic retrieval over source-code fragments;
  \item integration of LLM-derived quality metadata with vector-based retrieval;
  \item comparison of semantic, quality-filtered, hybrid, and automatically routed retrieval modes;
  \item analysis of observed limits, especially small query count, sample size, and quality-judge errors.
\end{itemize}

The evaluation is organized around four research questions:
\begin{description}
  \item[\textbf{RQ1.}] How effective are semantic, quality-filtered, hybrid, and automatic retrieval modes for natural-language queries over the evaluated source-code corpus?
  \item[\textbf{RQ2.}] Does explicit quality metadata improve retrieval for quality-oriented and mixed semantic-quality queries?
  \item[\textbf{RQ3.}] How accurately does the router select retrieval modes, and what latency does it add?
  \item[\textbf{RQ4.}] How reliable are the LLM-derived quality scores in the reported manual audit?
\end{description}

\section{Background}

Semantic code search is the task of retrieving code fragments from natural-language descriptions of the desired behavior. Husain et al. formalized this task through CodeSearchNet, a benchmark containing approximately six million functions in six programming languages paired with natural-language documentation \cite{Husain2019}. A common architecture for this task is the dual encoder: one encoder maps a query into a vector, another maps code into the same vector space, and cosine similarity is used to rank fragments. This design is attractive for search because all code vectors can be computed offline and stored in an index.

Several model families have been developed for this shared text-code space. CodeBERT extends BERT-style pretraining to paired natural-language and programming-language data \cite{Feng2020}. GraphCodeBERT adds data-flow information, showing that structural properties of programs can improve code representation beyond token sequences \cite{Guo2021}. The evaluated approach follows a more general route. It uses \tool{OpenAI} text-and-code embeddings, motivated by contrastive pretraining results showing that a single embedding model can map natural language and code into a shared semantic space \cite{Neelakantan2022}. This choice avoids training a code-specific model and supports practical deployment through an API.

\Rag{} combines retrieval from an external knowledge base with generative language modeling. Lewis et al. introduced the approach for knowledge-intensive natural-language processing tasks by pairing a sequence-to-sequence model with a dense retriever \cite{Lewis2020}. Later surveys emphasize three practical advantages: the knowledge base can be updated without retraining the model, answers can be linked to retrieved evidence, and the approach works with models available only through API access \cite{Gao2024}. These advantages are directly relevant for code repositories because the corpus is local, changes over time, and must be cited accurately.

In the present setting, RAG is not used to generate new code. It is used to retrieve existing source code and generate explanations grounded in that code. This distinction matters. Benchmarks such as CodeRAG-Bench focus on whether retrieval can improve code generation \cite{Wang2025coderag}. The present evaluation instead examines whether retrieval modes can surface relevant fragments and programmer identifiers from an existing educational C-code corpus. The generative model is therefore a reporting layer over retrieved evidence.

Dense semantic retrieval requires efficient search in high-dimensional vector spaces. Exact nearest-neighbor search compares the query vector against every indexed vector and scales linearly with the number of fragments. For larger corpora, approximate nearest-neighbor (ANN) indexes trade a small amount of exactness for faster retrieval. The system uses Hierarchical Navigable Small World (HNSW) indexing, which organizes vectors in a multi-layer graph and supports efficient approximate search \cite{Malkov2020}. The vectors and metadata are stored in \tool{ChromaDB}, an embedded vector database with Python support and metadata-aware collections \cite{ChromaDocs2024}.

The prototype adds a second signal besides vector similarity: an LLM-generated quality profile for every indexed fragment. The quality scorer follows the LLM-as-a-judge paradigm, where a language model is instructed to assign numeric ratings to code according to an explicit rubric. Related work such as ICE-Score evaluates code with LLM-based judgments and compares them with human assessment \cite{Zhuo2024}. In computer science education, automated feedback and LLM-assisted grading studies similarly emphasize that code quality signals require clear procedures and human oversight \cite{horvath2026personalized,horvath2026benchmarking}. The evaluated method applies this idea at indexing time, so that quality scores become metadata used later for filtering and re-ranking. The eight dimensions are edge-case handling, clean code, readability, efficiency, memory management, input validation, error handling, and documentation.

\subsection{Research gap}

Existing semantic-code-search and RAG evaluations usually emphasize broad benchmark corpora, code generation, or general software-engineering retrieval tasks. Less is known about retrieval behavior in small C-code repositories where many source-code submissions solve the same tasks and where queries may combine functional intent with quality-oriented criteria. In this setting, explicit quality metadata may help quality-centered questions, but it may also distort topical relevance. This motivates a direct comparison of semantic, quality-filtered, hybrid, and automatically routed retrieval rather than assuming that quality scores improve all search tasks.

\section{Dataset}

This section describes the educational C-code dataset and the fixed evaluation material used in the study. It separates the corpus and sampling procedure from the retrieval methods so that the scope of the reported results is clear.

\subsection{Evaluation material}

The evaluation uses a fixed set of recorded outputs from the retrieval prototype: the indexed sample, query set, pooled relevance judgments, router decisions, latency measurements, and manual quality-score audit. The software environment was a Python-based retrieval system configured with \tool{OpenAI} model calls and \tool{ChromaDB} vector storage. Implementation dependencies are treated only as reproducibility details, not as experimental variables.

\subsection{Corpus and sampling}

The full dataset contains 563 anonymized programmer identifiers and 8,951 C source files. The directory structure is assumed to encode metadata in the form \code{<year>/id\_N/psM/task\_K.c}, where \code{id\_N} is an anonymized programmer identifier, \code{psM} denotes a problem set, and \code{task\_K.c} denotes a concrete task. The evaluation used a reproducible 10\% sample created with seed 42. This sample contains 56 programmer identifiers, 847 files, and 3,839 indexed fragments in \tool{ChromaDB}.

The sample covers seven introductory C assignments: Karel-the-Robot navigation, array operations, Hangman, Ball Sort Puzzle, QR-code generation, Game of Life, and file processing. These assignments define the topical limits of the system. Queries outside this assignment coverage cannot be answered by retrieval, even if the embedding model can return semantically adjacent fragments.

Table \ref{tab:corpus} summarizes the corpus characteristics used in the reported evaluation.

\begin{table}[t]
\caption{Corpus characteristics used in the reported evaluation.}
\label{tab:corpus}
\centering
\begin{tabularx}{\linewidth}{@{}Y r@{}}
\toprule
\textbf{Property} & \textbf{Value} \\
\midrule
Programmer IDs in full corpus & 563 \\
C files in full corpus & 8,951 \\
Sampling ratio & 10\% \\
Random seed & 42 \\
Programmer IDs in indexed sample & 56 \\
Files in indexed sample & 847 \\
Indexed fragments & 3,839 \\
Dominant fragment type & functions (99.1\%) \\
Additional fragment types & 10 structs, 17 function parts, 7 files \\
\bottomrule
\end{tabularx}
\end{table}

\section{Methods}

The study follows a fixed retrieval methodology with four stages: corpus preprocessing, fragment-level indexing, retrieval-mode execution, and programmer-identifier-level evaluation. This section describes the methodological choices needed to interpret the experiment rather than the full software stack.

\subsection{Retrieval workflow}

\begin{figure*}[t]
\centering
\small
\begin{tikzpicture}[
  node distance=0.5cm and 0.25cm,
  proc/.style={draw, rounded corners=1pt, align=center, minimum height=0.70cm, text width=1.55cm},
  store/.style={draw, rounded corners=1pt, align=center, minimum height=0.82cm, text width=1.75cm, fill=black!8},
  arrow/.style={-{Latex[length=2mm]}, line width=0.35pt}
]
\node[proc] (files) {C source\\files};
\node[proc, right=of files] (parser) {Parser and\\fragmenter};
\node[proc, right=of parser] (quality) {Quality\\scorer};
\node[proc, right=of quality] (embed) {Embedding\\model};
\node[store, right=of embed] (db) {\emph{ChromaDB}\\vectors + metadata};
\node[proc, right=of db] (retrieval) {Retrieval\\mode};
\node[proc, right=of retrieval] (grouping) {Identifier\\grouping};
\node[proc, right=of grouping] (answer) {Optional\\grounded answer};
\draw[arrow] (files) -- (parser);
\draw[arrow] (parser) -- (quality);
\draw[arrow] (quality) -- (embed);
\draw[arrow] (embed) -- (db);
\draw[arrow] (db) -- (retrieval);
\draw[arrow] (retrieval) -- (grouping);
\draw[arrow] (grouping) -- (answer);
\node[align=center, below=0.45cm of db, text width=5.6cm] (query) {Natural-language query embedded and routed to semantic, quality-filtered, hybrid, or automatic retrieval};
\draw[arrow] (query.north) -- (db.south);
\end{tikzpicture}
\caption{Retrieval workflow used in the evaluation. C source files are parsed into fragments, scored, embedded, and stored with metadata in \tool{ChromaDB}. At query time, a retrieval mode returns fragments that are grouped by programmer identifier; generated answers are optional and are not separately evaluated in this study.}
\label{fig:pipeline}
\end{figure*}
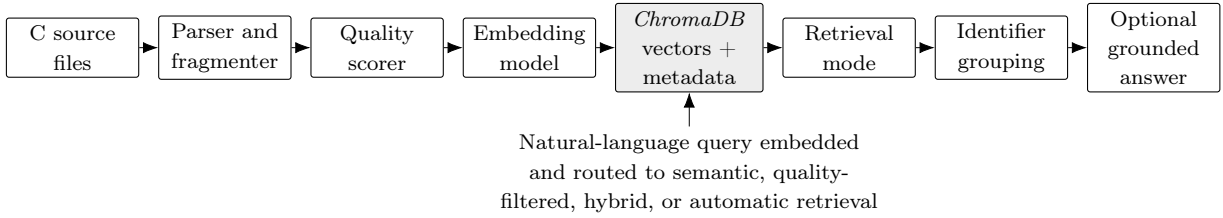

The retrieval workflow is organized as two linked phases. The offline phase converts source files into indexed fragments with quality metadata and vector representations. The online phase embeds a natural-language query, retrieves fragments, groups evidence by programmer identifier, and optionally generates a grounded answer from the retrieved evidence. Figure \ref{fig:pipeline} summarizes this workflow.

Table \ref{tab:components} summarizes how the main methodological stages are operationalized.

\begin{table*}[t]
\caption{Main methodological stages used in the retrieval evaluation.}
\label{tab:components}
\centering
\begin{tabularx}{\textwidth}{p{0.19\textwidth} p{0.23\textwidth} Y}
\toprule
\textbf{Stage} & \textbf{Operationalization} & \textbf{Purpose in the study} \\
\midrule
Corpus scanning & Identifier directories matching the \code{id\_N} pattern & Defines the population of anonymized source-code submissions used for sampling. \\
Fragment extraction & Function and struct extraction with lexer-aware brace counting & Defines the unit of indexing, scoring, retrieval, and judgment. \\
Quality scoring & Configured \tool{OpenAI} judge and eight quality dimensions & Produces metadata for quality-aware retrieval modes. \\
Embedding & \code{text-embedding-3-small} & Produces 1,536-dimensional vectors for fragment text and queries. \\
Vector storage & \tool{ChromaDB}, cosine HNSW & Stores vectors, code text, and metadata for nearest-neighbor retrieval. \\
Retrieval & semantic, filter, hybrid, auto & Provides the four retrieval modes compared in the evaluation. \\
Grounded response & Citation-based context from retrieved fragments & Optional reporting layer; generated answers are not separately evaluated. \\
\bottomrule
\end{tabularx}
\end{table*}

\subsection{Corpus preprocessing and indexing}

The indexing workflow begins by scanning the configured data directory and selecting identifier directories whose names match the \code{id\_N} convention. Sampling, when enabled, is applied at the programmer-identifier level before files are opened. The reported 10\% sample therefore represents 56 complete programmer identifiers rather than a random mixture of files from all identifiers.

After scanning, each non-empty source file is normalized before parsing. The parser then produces fragments with source text and line ranges. Oversized fragments are split before embedding or answer generation.

Duplicate removal is performed before quality scoring and embedding. The preprocessing procedure computes a SHA-256 hash over code after stripping leading and trailing whitespace from every non-empty line and removing blank lines. This exact-duplicate handling is narrower than clone detection, where related work distinguishes exact, renamed, modified, and semantic clones \cite{bubenkova2024clone}. The hash is reused as the \tool{ChromaDB} document identifier, making the record key deterministic.

For each non-duplicate fragment, the system obtains a quality profile, constructs embedding text, batches embedding calls, and upserts vectors and metadata into \tool{ChromaDB}.

\subsection{Fragment definition}

Fragmentation is central to RAG over source code because it determines the unit that is embedded and retrieved. The method uses function-level fragments whenever possible. A function is more semantically coherent than an arbitrary text window and small enough to fit within embedding limits. Struct definitions are extracted as separate fragments. If a file contains no recognizable function or struct, the entire file is indexed as one fragment.

The fragmenter must count braces without being misled by braces that occur inside strings, character literals, or block comments. The procedure therefore uses a small state machine with four states: normal code, string literal, character literal, and block comment. Line comments terminate processing of the current line, while block-comment state is carried across lines. This lets the parser track function and struct boundaries more reliably than a simple character count.

Function definitions are detected by two regular-expression patterns: one for a signature whose opening brace appears on the same line and one for a signature where the opening brace is on a following line. Control-flow keywords such as \code{if}, \code{while}, \code{for}, \code{switch}, \code{return}, and \code{else} are excluded to reduce false positives. Very small fragments with fewer than three non-empty lines are skipped. Oversized fragments above 12,000 characters are split into smaller parts while retaining parent-function metadata. Before indexing, code is normalized by replacing tabs with spaces, trimming trailing whitespace, capping excessive indentation, removing repeated blank lines, and computing a SHA-256 hash over whitespace-normalized code for deduplication.

\subsection{Quality metadata and vector representation}

Every fragment is graded by an LLM prompt that asks for an integer from 1 to 10 on each of eight dimensions. If a dimension is not applicable, the prompt instructs the model to return the neutral score 5. The response must be a JSON object and is clipped to the valid interval by the scoring procedure. If the API call fails or the JSON response is invalid, all dimensions default to 5 so that indexing can continue.

The resulting values are used as ranking metadata, not as ground-truth quality labels. This distinction is important because filter and hybrid retrieval depend on these metadata values, while the reported evaluation does not establish them as calibrated assessments of code quality.

Table \ref{tab:quality-dimensions} defines the eight quality dimensions stored as fragment metadata.

\begin{table*}[t]
\caption{Quality dimensions used as metadata for each indexed code fragment.}
\label{tab:quality-dimensions}
\centering
\begin{tabularx}{\textwidth}{p{0.24\textwidth} Y}
\toprule
\textbf{Dimension} & \textbf{Operational meaning} \\
\midrule
\code{edge\_case\_handling} & Null checks, boundary checks, empty-input guards, and checks of error return values. \\
\code{clean\_code} & Naming conventions, formatting consistency, and logical organization. \\
\code{readability} & Meaningful variable and function names, clear structure, and suitable comments. \\
\code{efficiency} & Appropriate algorithm choice, absence of unnecessary computation, and suitable data-structure use. \\
\code{memory\_management} & Correct \code{malloc}/\code{free} pairing, absence of leaks, and no use-after-free patterns. \\
\code{input\_validation} & Checks of parameter and input validity before processing. \\
\code{error\_handling} & Proper reaction to error states, error codes, and informative messages. \\
\code{documentation} & Presence and quality of comments describing purpose and behavior. \\
\bottomrule
\end{tabularx}
\end{table*}

The grading function is intentionally executed at indexing time rather than query time. This design makes quality-oriented search cheap during interaction: quality values are ordinary metadata fields in \tool{ChromaDB}, not fresh model calls. The trade-off is that indexing requires one model call per non-duplicate fragment, and errors in the LLM judge become persistent metadata until the database is rebuilt.

After quality scoring, each fragment is converted into an embedding input containing file path, fragment type, fragment name, language, and source code. The quality scores are not included in the embedding text in the reported configuration; they are stored only as metadata. This separates semantic similarity from the explicit quality signal. The embedding call uses \code{text-embedding-3-small}, which produces 1,536-dimensional vectors \cite{OpenAIEmbeddings2024}. Text is truncated to at most 5,000 tokens before embedding.

The \tool{ChromaDB} collection is created as \code{code\_snippets} with cosine distance metadata for HNSW indexing. For each stored record, the document text is the fragment code, the vector is the embedding, and the metadata contains file path, programmer identifier, fragment type, fragment name, language, line range, overall quality, and the eight individual quality dimensions. The SHA-256 hash of normalized code is used as the record identifier.

\subsection{Retrieval methods}

The evaluation compares four retrieval methods. The first three are direct methods, and the fourth delegates method choice to an LLM router.

\textbf{Semantic mode.} The query is embedded with the same model used during indexing. \tool{ChromaDB} returns the nearest fragments by cosine distance. The method converts a returned distance $d$ to a similarity score on a 0--10 scale:
\begin{equation}
S = \max(0,\min(1,1-d/2)) \cdot 10 .
\label{eq:similarity}
\end{equation}

\textbf{Filter mode.} The mode targets quality-centered questions. It first retrieves a larger semantic fragment pool of size $\min(100,\max(50,3n))$, where $n$ is the requested number of results. Each fragment receives a quality score
\begin{equation}
Q = \frac{1}{|D|}\sum_{d \in D} q_d ,
\label{eq:quality}
\end{equation}
where $D$ is the active set of quality dimensions and $q_d$ is the stored score for dimension $d$. If forced filter mode receives no dimensions, all eight dimensions are used. Fragments are sorted by descending $Q$ and then by cosine distance.

\textbf{Hybrid mode.} The mode retrieves a broader semantic pool, 120 by default and capped at 200, and combines quality and similarity:
\begin{equation}
H = \alpha Q + (1-\alpha)S,\quad \alpha = 0.5 .
\label{eq:hybrid}
\end{equation}
The default assigns equal weight to quality and similarity. The method stores both \code{quality\_score} and \code{combined\_score}, sorts by the combined score, and keeps the top $n$ fragments.

\textbf{Automatic mode.} The router sends the original query to the configured \tool{OpenAI} routing model with minimal reasoning effort and three tool definitions: \code{semantic\_search}, \code{filter\_search}, and \code{hybrid\_search}. For filter and hybrid tools, the model must select quality dimensions from the fixed eight-item enumeration. If the router fails, returns no tool call, or returns an unknown tool, the system falls back to semantic search. If all returned dimensions are invalid, it substitutes all eight dimensions.

\subsection{Model configuration and answer-generation constraints}

The method uses separate configured models for embedding, quality scoring, routing, and optional answer generation. The embedding model is \code{text-embedding-3-small}; the remaining model calls are described by their component roles because API model identifiers can change over time. Fragment embedding text is truncated to 5,000 tokens. Answer generation receives at most three ranked programmer identifiers and at most three fragments for each identifier, giving a maximum of nine code excerpts.

At response time, retrieved fragments are grouped by programmer identifier. Fragment relevance is computed as semantic similarity plus any mode-specific quality score available for filter or hybrid results. For each programmer identifier, fragments are sorted by relevance, and the identifier score is the mean relevance of the top five fragments. This cap reduces the advantage of identifiers with many indexed fragments.

For answer generation, each excerpt includes a citation tag with programmer identifier, file path, function name, and line range. The system prompt requires factual claims about a programmer identifier to cite an excerpt, requires absent features to be reported as not found in retrieved code, and forbids invented file paths, function names, libraries, errors, or behaviors. The optional verify pass removes sentences lacking excerpt citations or misrepresenting the provided context.

\section{Evaluation Protocol}

The evaluation compares retrieval modes rather than generated prose answers. It uses a fixed query set, pooled judgments, and standard ranking metrics to measure whether each mode surfaces relevant programmer identifiers in the top results. The evaluation reported here assesses retrieval effectiveness and supporting components. It does not provide a separate faithfulness evaluation of generated natural-language answers.

\subsection{Query set and relevance judgments}

The evaluation used 15 English-language test queries divided into three categories: five semantic queries, five quality queries, and five hybrid queries. Each query was executed in all four modes with \code{n\_results=30} and \code{top\_candidates=5}, giving 60 retrieval operations. Programmer identifiers returned by the modes were pooled per query. Each pooled programmer identifier was then manually judged on a 0--3 relevance scale by reading the retrieved code:
\begin{itemize}
  \item 0: not relevant;
  \item 1: marginally relevant;
  \item 2: relevant;
  \item 3: highly relevant.
\end{itemize}

Table \ref{tab:all-queries} lists the complete query set grouped by expected retrieval mode. The reported metrics are nDCG@5, Success@$k$, mean reciprocal rank (MRR), and Jaccard overlap of returned programmer identifiers. nDCG@5 rewards both high relevance and high rank in the top five results. Success@$k$ records whether at least one programmer identifier with relevance at least 2 appears in the first $k$ results. MRR measures how early the first relevant programmer identifier appears. Jaccard overlap measures how similar two modes are in terms of returned programmer identifiers.

\begin{table*}[t]
\caption{Complete test-query set grouped by expected retrieval mode.}
\label{tab:all-queries}
\centering
\begin{tabularx}{\textwidth}{p{0.18\textwidth} Y}
\toprule
\textbf{Expected mode} & \textbf{Queries} \\
\midrule
semantic & \textit{Find linked list implementations}; \textit{Show sorting algorithm solutions}; \textit{Find recursive function implementations}; \textit{Show implementations that read from files}; \textit{Find code that uses dynamic memory allocation with malloc}. \\
filter & \textit{Who writes the cleanest code?}; \textit{Which students best handle edge cases?}; \textit{Find code with the best error handling}; \textit{Who has the worst input validation?}; \textit{Which students write the most readable code?}. \\
hybrid & \textit{Clean implementation of dynamic memory allocation}; \textit{Robust file reading with proper error handling}; \textit{Efficient and readable array operations}; \textit{Well-documented sorting implementation}; \textit{Safe string handling with good memory management}. \\
\bottomrule
\end{tabularx}
\end{table*}

\subsection{Metric computation}

For completeness, the metrics can be written directly from the relevance list $r_1,\ldots,r_k$ for a ranked list of programmer identifiers. Discounted cumulative gain at rank $k$ is
\begin{equation}
\mathrm{DCG@}k = \sum_{i=1}^{k}\frac{r_i}{\log_2(i+1)} .
\end{equation}
The normalized score divides this value by the ideal DCG for the same judged pool. Success@$k$ is 1 when at least one of the first $k$ relevance values is 2 or 3, and 0 otherwise. MRR is the reciprocal rank of the first result whose relevance is at least 2. The Jaccard overlap between two modes $A$ and $B$ is
\begin{equation}
J(A,B) = \frac{|A\cap B|}{|A\cup B|},
\end{equation}
where $A$ and $B$ are sets of returned programmer identifiers for the same query.

\section{Results}

The results are grouped into four parts: indexed-corpus characteristics, retrieval performance, router behavior with latency, and a manual audit of the quality-scoring component.

\subsection{Indexed corpus and quality scores}

The 10\% sample produced 3,839 indexed fragments. Functions dominate the sample, accounting for 99.1\% of all fragments. The remaining fragments are 10 struct definitions, 17 split function parts, and 7 whole-file fragments. The reported retrieval results therefore primarily concern function-level fragments.

The indexed sample covers seven assignment families. The linked-list query functions as an out-of-coverage case: the retrieval system is forced to return nearest neighbours even when the requested concept is absent from the indexed assignments. Because none of the seven assignments required linked-list implementations, all modes returned irrelevant fragments for that query and achieved nDCG@5 of 0.

Table \ref{tab:quality-distribution} reports the distribution of automatically assigned quality scores. Documentation has the lowest mean score, 2.23, followed by edge-case handling at 2.69 and error handling at 2.86. The highest means are memory management at 5.38 and efficiency at 5.15. Because the prompt defines 5 as the neutral value for inapplicable dimensions, mid-range values may also reflect fragments where the dimension was not applicable.

\begin{table}[t]
\caption{Distribution of automatically assigned quality scores across 3,839 fragments.}
\label{tab:quality-distribution}
\centering
\begin{tabular}{lcc}
\toprule
\textbf{Dimension} & \textbf{Mean} & \textbf{SD} \\
\midrule
\code{edge\_case\_handling} & 2.69 & 1.10 \\
\code{clean\_code} & 5.09 & 1.46 \\
\code{readability} & 5.12 & 1.60 \\
\code{efficiency} & 5.15 & 1.76 \\
\code{memory\_management} & 5.38 & 1.36 \\
\code{input\_validation} & 3.16 & 1.49 \\
\code{error\_handling} & 2.86 & 1.23 \\
\code{documentation} & 2.23 & 0.73 \\
\midrule
\code{overall\_quality} & 3.96 & 0.86 \\
\bottomrule
\end{tabular}
\end{table}

The later manual audit shows that trivial functions can be mis-scored when a dimension should be treated as not applicable.

\subsection{Retrieval performance across modes}

Table \ref{tab:retrieval} presents the main retrieval metrics averaged over the 15 queries. Semantic mode has the highest aggregate nDCG@5, 0.820, with Success@1 of 0.533, Success@3 and Success@5 of 0.800, and MRR of 0.644. The filter, hybrid, and auto modes have aggregate nDCG@5 values of 0.668, 0.658, and 0.640, respectively.

\begin{table}[t]
\caption{Retrieval metrics averaged across 15 test queries.}
\label{tab:retrieval}
\centering
\begin{tabular}{lccccc}
\toprule
\textbf{Mode} & \textbf{nDCG@5} & \textbf{S@1} & \textbf{S@3} & \textbf{S@5} & \textbf{MRR} \\
\midrule
\code{semantic} & \textbf{0.820} & \textbf{0.533} & \textbf{0.800} & \textbf{0.800} & \textbf{0.644} \\
\code{filter} & 0.668 & 0.467 & 0.600 & 0.667 & 0.539 \\
\code{hybrid} & 0.658 & 0.400 & 0.600 & 0.600 & 0.489 \\
\code{auto} & 0.640 & 0.467 & 0.600 & 0.667 & 0.539 \\
\bottomrule
\end{tabular}
\end{table}

The category breakdown in Table \ref{tab:category} reveals a more nuanced pattern. On semantic queries, the semantic mode has the highest nDCG@5 at 0.764. Filter and hybrid modes are lower, 0.324 and 0.401, because quality re-ranking can move topically relevant fragments below higher-quality but off-topic fragments.

\begin{table}[t]
\caption{nDCG@5 by query category. Each category contains five queries.}
\label{tab:category}
\centering
\begin{tabular}{lccc}
\toprule
\textbf{Mode} & \textbf{Semantic} & \textbf{Quality} & \textbf{Hybrid} \\
\midrule
\code{semantic} & \textbf{0.764} & 0.939 & \textbf{0.757} \\
\code{filter} & 0.324 & 0.947 & 0.733 \\
\code{hybrid} & 0.401 & \textbf{0.959} & 0.613 \\
\code{auto} & 0.540 & 0.898 & 0.481 \\
\bottomrule
\end{tabular}
\end{table}

On quality queries, hybrid mode is highest at 0.959, filter mode follows at 0.947, semantic mode reaches 0.939, and auto mode reaches 0.898. On hybrid queries, semantic mode leads with 0.757, followed by filter at 0.733 and hybrid at 0.613.

Returned-identifier overlap gives a second view of mode behavior. The average Jaccard overlap between semantic and filter results is 0.106, and between semantic and hybrid results it is 0.061. In contrast, filter and hybrid overlap at 0.482. Table \ref{tab:overlap} reports the full pairwise overlap values.

\begin{table}[t]
\caption{Average Jaccard overlap of returned programmer identifiers between retrieval modes.}
\label{tab:overlap}
\centering
\begin{tabular}{lc}
\toprule
\textbf{Mode pair} & \textbf{Jaccard overlap} \\
\midrule
\code{semantic} vs. \code{filter} & 0.106 \\
\code{semantic} vs. \code{hybrid} & 0.061 \\
\code{semantic} vs. \code{auto} & 0.112 \\
\code{filter} vs. \code{hybrid} & 0.482 \\
\code{filter} vs. \code{auto} & 0.259 \\
\code{hybrid} vs. \code{auto} & 0.236 \\
\bottomrule
\end{tabular}
\end{table}

The low overlap indicates that quality-aware retrieval changes which programmer identifiers appear in the top-ranked results, not only the order of a shared result set.

\subsection{Router accuracy and latency}

The automatic router classified all 15 test queries into the expected mode, giving 15/15 correct decisions. It routed the five semantic queries to semantic search, the five quality queries to filter search, and the five hybrid queries to hybrid search. For filter and hybrid queries it also selected relevant quality dimensions from the fixed set. For the edge-case handling query, for example, it selected edge-case handling, readability, and error handling.

\begin{figure*}[t]
\centering
\small
\begin{tikzpicture}[x=1cm,y=1cm]
  \node[anchor=west,font=\bfseries] at (0,3.75) {(a) Mean retrieval latency (s)};
  \draw[->] (0,0) -- (0,3.25);
  \draw[->] (0,0) -- (5.0,0);
  \foreach \y/\tick in {0/0,1.5/1,3.0/2} {
    \draw (-0.05,\y) -- (0,\y);
    \node[anchor=east] at (-0.10,\y) {\tick};
  }
  \fill[black!45] (0.45,0) rectangle (1.00,0.272);
  \node[anchor=south] at (0.725,0.272) {0.181};
  \node[anchor=north] at (0.725,-0.12) {\code{semantic}};
  \fill[black!45] (1.55,0) rectangle (2.10,0.210);
  \node[anchor=south] at (1.825,0.210) {0.140};
  \node[anchor=north] at (1.825,-0.12) {\code{filter}};
  \fill[black!45] (2.65,0) rectangle (3.20,0.216);
  \node[anchor=south] at (2.925,0.216) {0.144};
  \node[anchor=north] at (2.925,-0.12) {\code{hybrid}};
  \fill[black!75] (3.75,0) rectangle (4.30,2.865);
  \node[anchor=south] at (4.025,2.865) {1.910};
  \node[anchor=north] at (4.025,-0.12) {\code{auto}};

  \node[anchor=west,font=\bfseries] at (6.1,3.75) {(b) Correct router decisions};
  \draw[->] (6.1,0) -- (6.1,3.25);
  \draw[->] (6.1,0) -- (10.8,0);
  \foreach \y/\tick in {0/0,1.5/2.5,3.0/5} {
    \draw (6.05,\y) -- (6.1,\y);
    \node[anchor=east] at (6.00,\y) {\tick};
  }
  \fill[black!65] (6.75,0) rectangle (7.30,3.00);
  \node[anchor=south] at (7.025,3.00) {5/5};
  \node[anchor=north] at (7.025,-0.12) {\code{semantic}};
  \fill[black!65] (8.05,0) rectangle (8.60,3.00);
  \node[anchor=south] at (8.325,3.00) {5/5};
  \node[anchor=north] at (8.325,-0.12) {\code{filter}};
  \fill[black!65] (9.35,0) rectangle (9.90,3.00);
  \node[anchor=south] at (9.625,3.00) {5/5};
  \node[anchor=north] at (9.625,-0.12) {\code{hybrid}};
\end{tikzpicture}
\caption{Latency and router behavior. Direct retrieval modes were below 0.2 s on average. Auto mode includes an additional router call. Router correctness is based on 15 predefined queries, five per category.}
\label{fig:router-latency}
\end{figure*}
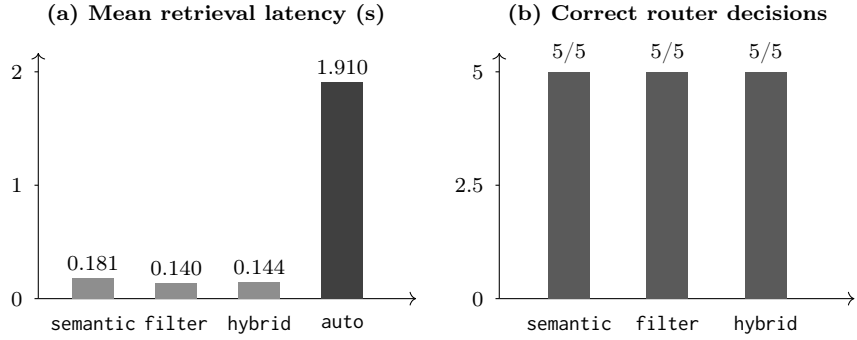

Figure \ref{fig:router-latency} summarizes the measured latency and router correctness. Direct retrieval modes were below 0.2 s on average: 0.181 s for semantic, 0.140 s for filter, and 0.144 s for hybrid. Auto mode averaged 1.910 s, with a median of 1.579 s, minimum of 1.205 s, and maximum of 5.913 s. The additional cost comes from the router model call.

The router's 15/15 expected-mode match must be interpreted with caution because the query set is small and the expected mode labels were constructed by the evaluator. Within this query set, constrained tool calling mapped the predefined natural-language code-search questions to the available retrieval modes.

\subsection{Manual quality-score audit and qualitative retrieval cases}

The quality scores assigned by the configured quality-judge model were manually checked on a random sample of 12 fragments from different assignments and programmer identifiers. This manual audit is a small diagnostic check, not a reliability or calibration study. The manual assessment matched the automatic score within $\pm 1$ point in 9 of the 12 cases. Table \ref{tab:quality-audit} lists representative audit cases.

\begin{table*}[t]
\caption{Representative manual quality-score audit cases. Dimension abbreviations: eh = edge-case handling, iv = input validation, rd = readability, ef = efficiency, cc = clean code.}
\label{tab:quality-audit}
\centering
\begin{tabularx}{\textwidth}{p{0.24\textwidth} p{0.08\textwidth} p{0.08\textwidth} Y}
\toprule
\textbf{Fragment} & \textbf{Dim.} & \textbf{LLM} & \textbf{Manual assessment} \\
\midrule
\code{read\_file()} from \code{id\_36/ps7} & eh & 2 & Correct: missing \code{fopen() == NULL} check. \\
\code{array\_min()} from \code{id\_311/ps2} & iv & 1 & Correct: missing null-pointer check. \\
\code{check()} from \code{id\_587/ps1} & rd & 2 & Correct: dense code without spaces and comments. \\
\code{end()} from \code{id\_422/ps1} & ef & 9 & Overrated: trivial function; score 5 would be more appropriate. \\
\code{main()} from \code{id\_225/ps5} & ef & 10 & Overrated: empty function; dimensions are not applicable. \\
\code{compare()} from \code{id\_107/ps3} & cc & 7 & Adequate: clean one-line comparison function for \code{qsort}. \\
\bottomrule
\end{tabularx}
\end{table*}

The three disagreements occurred on trivial or empty functions. In these cases, the judge sometimes assigned a high score where the prompt required the neutral value 5 for non-applicable dimensions. This is a concrete failure mode of the LLM-as-a-judge component and can affect filter and hybrid retrieval when such fragments enter the result pool.

Two qualitative examples clarify the quantitative results. For the in-domain query ``Show implementations that read from files'', semantic search returned fragments containing calls such as \code{fopen()}, \code{fgetc()}, and \code{fclose()}. The top three programmer identifiers were manually rated as highly relevant. Filter mode shared two of the top five programmer identifiers with semantic mode but also promoted high-quality fragments that operated on arrays rather than files. Hybrid mode returned mostly code from the QR-code assignment, which was high quality but not directly related to file reading.

The out-of-domain query ``Find linked list implementations'' behaved differently. The indexed assignments did not include linked lists. All modes returned fragments from Karel or utility functions, and all achieved nDCG@5 of 0. This illustrates the retrieval-bounded nature of the system: a RAG pipeline can ground answers in its index, but it cannot retrieve content that is absent from that index.

\section{Discussion}

The results should be interpreted as evidence about the evaluated prototype and query set, not as a general claim about source-code repositories. The main findings concern the relative behavior of semantic retrieval, quality-aware retrieval, automatic routing, and the LLM-derived quality metadata in this educational C-code repository.

The research questions can be answered directly from the reported evaluation:
\begin{description}
  \item[\textbf{RQ1.}] Semantic retrieval was the strongest aggregate mode in the evaluated corpus, with nDCG@5 of 0.820, Success@5 of 0.800, and MRR of 0.644 across the 15 judged queries.
  \item[\textbf{RQ2.}] Quality-aware modes provided the highest nDCG@5 for explicitly quality-oriented queries, where hybrid and filter modes reached 0.959 and 0.947 respectively, but they did not improve the overall result compared with pure semantic retrieval.
  \item[\textbf{RQ3.}] The automatic router selected the expected mode for all 15 queries, but the auto mode averaged 1.910 s, while the direct modes averaged below 0.2 s.
  \item[\textbf{RQ4.}] In the manual audit, the quality judge matched manual assessment within one point for 9 of 12 fragments; the documented errors were concentrated in trivial or empty functions.
\end{description}

\subsection{Main interpretation and practical implications}

The strongest aggregate result belongs to pure semantic retrieval. This does not show that explicit quality metadata is unnecessary, but it does show that quality-aware re-ranking did not improve overall retrieval in the reported query set. One possible explanation is that the embedding representation already aligns with some quality-related query terms, but this remains an interpretation rather than a demonstrated causal mechanism.

Quality-aware retrieval still provides a distinct view of the corpus. The low Jaccard overlap between semantic and quality-aware results shows that the modes return different programmer identifiers. This can be useful for exploratory inspection, but it also creates a risk: quality scores can promote fragments that are cleaner or better documented while being less topically relevant to the query. This trade-off is visible in the semantic-query category, where filter and hybrid modes perform worse than semantic retrieval.

In the evaluated setting, the prototype is best interpreted as an exploratory retrieval tool. It supports questions that would be tedious to answer by manual inspection, such as which programmer identifiers are associated with file I/O, error checks, readable implementations, or missing edge-case handling. Because results are grouped by programmer identifier and linked to source excerpts, they can guide closer review. The reported evaluation does not support use as an automated grading or assessment system.

The linked-list query illustrates the corpus-bounded nature of retrieval. When the requested concept is absent from the indexed assignments, the system still returns nearest neighbours, but those neighbours need not be relevant. A user-facing system should therefore distinguish between low confidence, absent corpus coverage, and an ambiguous query.

\subsection{Grounding and hallucination control}

The answer-generation layer is deliberately constrained. It supplies retrieved excerpts and requires excerpt citations for factual claims. It also includes a verify pass that removes unsupported statements. These design choices address the known risk of hallucination in LLM outputs \cite{Huang2025}. However, the evaluation did not separately measure final-answer faithfulness. Retrieval metrics and router accuracy show whether relevant fragments were surfaced; they do not show that every generated sentence is correct.

\subsection{Threats to validity}

The first limitation is sample size. The experiment uses a 10\% sample, 56 of 563 programmer identifiers, and 15 queries. Fifteen queries are insufficient for stable estimates or statistical significance testing, and the study does not include a full-corpus evaluation. The within-sample comparison of modes is useful, but conclusions should not be generalized to the full corpus or other source-code repositories without repeating the evaluation on larger or multiple samples.

The second limitation is the annotation process. The manuscript does not document multiple relevance annotators or inter-annotator agreement. Because judgments were pooled from retrieved programmer identifiers, unreturned but relevant identifiers were not assessed. This pooled-judgment design is common in retrieval evaluation but can bias estimates when retrieval modes miss relevant material.

The third limitation is model specificity. All reported retrieval results use a single embedding model, \code{text-embedding-3-small}. A code-specialized model such as GraphCodeBERT could change the relative ranking of modes \cite{Guo2021}. Similarly, the quality scorer and router use specific configured \tool{OpenAI} models, including a single quality-judge model. Different models, prompts, or temperature settings could produce different quality metadata and routing behavior.

The fourth limitation is the quality-scoring audit. Twelve fragments are insufficient to estimate reliability, calibration, or inter-rater agreement. The observed errors are also systematic: trivial or empty functions can be overrated in dimensions that should be neutral or not applicable. Because filter and hybrid modes depend on these scores, scoring errors and quality-aware re-ranking can distort topical relevance. Quality-oriented query wording may also overlap with the wording used in LLM-derived quality labels, which could affect the relation between query text and stored metadata.

The fifth limitation is corpus coverage and answer evaluation. The corpus is a local educational C-code repository with assignment-specific coverage. The linked-list query illustrates that absent concepts cannot be recovered by retrieval. In addition, the evaluation does not include an answer-level faithfulness assessment of generated natural-language responses.

\subsection{Scalability, responsible use, and further evaluation}

The dominant indexing costs are LLM-based quality grading and embedding generation. For an illustrative dataset of 3,000 fragments, grading was estimated at about 0.10--0.15 USD under the API prices used during implementation. Embedding the same 3,000 fragments at an average length of roughly 400 tokens was estimated at about 0.02 USD. These values are implementation-time estimates and should not be interpreted as current API prices.

Deduplication removed approximately 10--20\% of fragments in the development dataset, and 3,000 records occupied roughly 20--30 MB, including 1,536-dimensional float32 vectors, source code strings, and metadata dictionaries. These values describe the documented implementation setting and should not be generalized without additional measurement.

Interactive retrieval was fast for direct modes in the measured experiment. Semantic, filter, and hybrid retrieval all averaged below 0.2 s. The auto mode was slower because it makes one additional router call. Answer generation and the optional verify pass add further model calls and were not included in the retrieval-effectiveness metrics.

Responsible use is a practical requirement. The system works with source-code submissions and produces quality-oriented rankings by programmer identifier. Even though identifiers are anonymized in the dataset, the quality scores are automatic model judgments, not final grades. The manual audit shows that the judge can be wrong, especially on trivial functions. Results should therefore guide inspection rather than support consequential decisions without human review.

The design helps with this by preserving evidence. Retrieved results include file paths, function names, line ranges, and source excerpts. This evidence-first workflow is preferable to presenting a bare ranking because it keeps the human reviewer connected to the actual code.

Privacy is a second concern. The manuscript does not describe public release of the evaluation corpus. Source code can contain stylistic and structural signals relevant to programmer profiling or attribution \cite{horvath2026bridging}. If such a system were deployed beyond a local controlled environment, data protection and institutional approval would be necessary. Queries and retrieved code sent to external model APIs may also carry privacy implications.

Further evaluation should repeat the experiment on the full 563-identifier corpus and expand the query set to at least 30--50 queries. Multiple human annotators should judge programmer-identifier relevance and quality scores to measure inter-annotator agreement.

A second direction is model comparison. The current evaluation should be repeated with at least one code-specific embedding model and one general embedding model, keeping all other components fixed. This would test whether the semantic baseline is a property of the selected model or of the corpus and query design. The hybrid formula should also be tuned rather than fixed at $\alpha=0.5$.

A third direction is improving the quality-scoring protocol. The observed ambiguity around neutral scores suggests an applicability-first procedure: first classify whether a dimension applies, then assign a 1--10 score only when it applies, and store ``not applicable'' separately rather than encoding it as 5. A fourth direction is answer-level evaluation. Generated responses should be assessed for citation accuracy, completeness, and unsupported claims, and the optional verify pass should be compared against generation without verification in terms of both faithfulness and latency.

\section{Conclusion}

This study evaluated a RAG-based retrieval prototype for semantic and quality-aware retrieval over source-code fragments. The concrete evaluation used an educational C-code corpus. The system fragments code at function boundaries, assigns LLM-derived quality metadata, embeds fragments with \code{text-embedding-3-small}, and stores vectors and metadata in \tool{ChromaDB}. The evaluation compared semantic, quality-filtered, hybrid, and automatically routed retrieval modes on a 10\% sample containing 56 programmer identifiers, 847 files, and 3,839 fragments.

In this evaluation, semantic retrieval was the strongest default mode, with nDCG@5 of 0.820, Success@5 of 0.800, and MRR of 0.644 across 15 manually judged queries. Explicit quality metadata provided an alternative ranking mainly for quality-oriented queries, but quality-aware modes did not improve the aggregate result. The router selected the expected mode for all 15 queries, but it added latency relative to direct retrieval. The manual quality-score audit found agreement within one point for 9 of 12 fragments and exposed errors on trivial or empty functions.

The results support the prototype as an exploratory retrieval tool for source-code repositories, with the reported evidence limited to the educational C-code corpus evaluated here. It should not be treated as an automated grading or assessment system. Stronger claims require a larger query set, full-corpus evaluation, multiple annotators, model comparisons, improved handling of non-applicable quality dimensions, and a separate faithfulness evaluation of generated answers.

\section*{Acknowledgement}

This work was supported by project VEGA No. 1/0708/26 ``Bridging the Understanding Gap between Testers and Programmers''.

\section*{Declaration on AI-Assisted Tools}

During preparation of this manuscript, \tool{ChatGPT} and \tool{Grammarly} were used for language polishing, grammar correction, LaTeX support, and formatting already computed numerical results into tables. These tools were not used to generate experimental results, fabricate data, or perform the reported evaluation. The authors take full responsibility for the content of the manuscript.

\bibliographystyle{plainnat}
\bibliography{references}

\end{document}